\documentclass{aa} 
\usepackage{times,amssymb,amsmath,epsfig,graphicx}

\newcommand{\chandra}{\textit{Chandra\/} }

\title{X-ray and weak lensing measurements of the mass profile of
  MS1008.1$-$1224: \textit{Chandra\/} and VLT data}

\author{S. Ettori \inst{1} \and M. Lombardi \inst{1,2}}

\institute{European Southern Observatory,  
           Karl-Schwarzschild-Stra\ss e 2,
           85748, Garching bei M\"unchen, Germany
           \and
           Instit\"ut f\"ur Astrophysik und Extraterrestrische Forschung,
           Auf dem H\"ugel 71, 53121, Bonn, Germany}
\offprints{}
\mail{\tt settori@eso.org}

\date{Accepted on 5th December 2002}

\begin{document}
\titlerunning{Mass profile in MS1008.1$-$1224}

\abstract{We analyse the \chandra dataset of the galaxy cluster
  MS1008.1$-$1224 to recover an estimate of the gravitating mass as
  function of the radius and compare these results with the weak
  lensing reconstruction of the mass distribution obtained from deep
  FORS1-VLT multicolor imaging.  Even though the X-ray morphology is
  disturbed with a significant excess in the northern direction
  suggesting that the cluster is not in a relaxed state, we are able
  to match the two mass profiles both in absolute value and in shape
  within $1 \sigma$ uncertainty and up to 1100 $h_{50}^{-1}$ kpc.
  The recovered X-ray mass estimate does not change by using either
  the azimuthally averaged gas density and temperature profiles or
  the results obtained in the northern sector alone where the
  signal-to-noise ratio is higher.
  }

\maketitle

\section{Introduction}

As the largest virialized objects in the Universe, galaxy clusters are
a powerful cosmological tool once their mass distribution is
univocally determined.  In the recent past, there have been several
claims that cluster masses obtained from X-ray analyses of the
intracluster plasma, taken to be in hydrostatic equilibrium with the
gravitational potential well, are significantly smaller (up to a
factor of two; but see Wu et al.\ 1998; Allen 1998; B\"ohringer et al.\ 
1998; Allen et al.\ 2001) than the ones derived from gravitational
lensing (see Mellier 1999 for a review).

\begin{figure}
  \begin{center}
    \epsfig{file=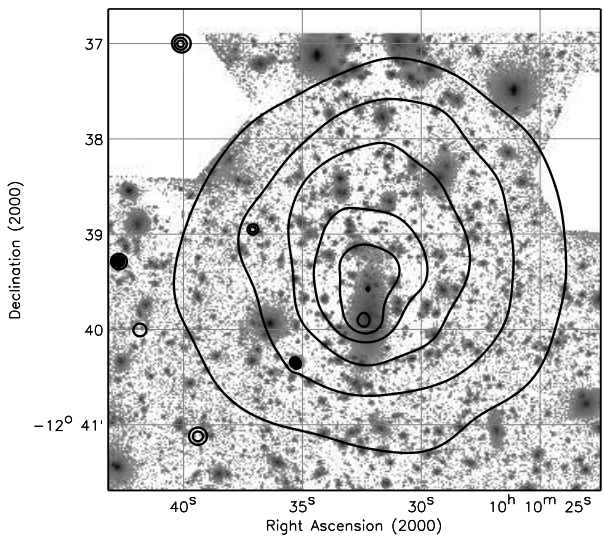,width=0.45\textwidth}
    \caption{An adaptive smoothed exposure-corrected image in the
      Chandra [0.5--2] keV band is overplotted to the R-band image (the white
areas in the upper end show masked regions). The
      contours are spaced with steps of $\log 2$ from the minimum of
      $2.1 \times 10^{-9} \mbox{ photon sec}^{-1} \mbox{ cm}^{-2}
      \mbox{ arcsec}^{-2}$.}
  \end{center}
  \label{opt_xray}
\end{figure}

In this paper we report on the mass distribution of the cluster
MS1008.1$-$1224 by combining the results from weak lensing analysis of deep
FORS1-VLT images with those obtained from a spatially-resolved
spectroscopic X-ray analysis of a \chandra observation.
MS1008.1$-$1224 is a rich galaxy cluster at redshift 0.302 that has been
part of the Einstein Medium Sensitivity Survey sample (Gioia \& Luppino
1994) and of the CNOC survey (Carlberg et al.\ 1996).  Lombardi et
al.\ (2000) presented a detailed weak lensing analysis of the FORS1-VLT
data.  Figure~\ref{opt_xray} shows the X-ray isophotes overplotted to
the optical V-band image.
In the following we adopt the conversion $1 \mbox{ arcmin} =
376 \mbox{ kpc}$ ($z=0.302$, $H_0 = 50 \, h_{50} \mbox{ km s}^{-1}
\mbox{ Mpc}^{-1}$, $\Omega_{\rm m} = 1 - \Omega_{\Lambda} = 0.3$) and
quote all the errors at $1 \sigma$ ($68.3\%$ confidence level).

\section{X-ray mass}

\begin{figure}
  \begin{center}
    \epsfig{file=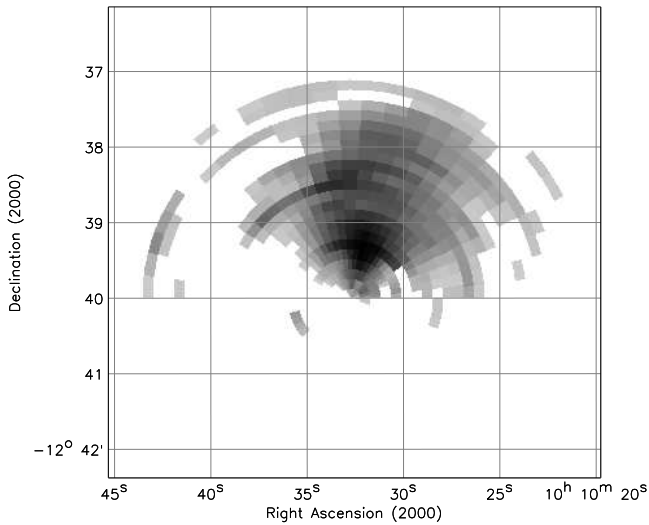,width=0.45\textwidth}
    \caption{Residuals above $2$ in the X-ray brightness
distribution. Given two regions, $A$ and $B$, of (length, width)${}
= (8$ arcsec$, 20^\circ)$ located symmetrically with respect to the center and
with $C_A$ and $C_B$ observed counts respectively, the residuals
are estimated as $\sigma = (C_A - C_B) / \sqrt{C_A + C_B}$. 
An azimuthal scan with step of $20^\circ/3$ was done to smooth the map.
The most significant excess is
      along $63^\circ \pm 10^\circ$ (anticlockwise, from X-axis) with a
      power of 0.87 estimated as the fraction of regions with higher
      excess along a fixed direction.} 
  \end{center}
  \label{xray_asi}
\end{figure}

We retrieved the primary and secondary data products from the \chandra
archive. The exposure of MS1008.1$-$1224 was done on June 11, 2000
using the ACIS-I configuration.  We reprocessed the level=1
events file in the Very Faint Mode and, then, with the
\textit{CtiCorrector\/} software (v.\ 1.38; Townsley et al.\ 2000).
The light curve was checked for high background flares that were
not detected.  About $44.0 \mbox{ ksec}$ (out of $44.2 \mbox{ ksec}$,
the nominal exposure time) were used and a total number of
counts of about $20 \, 000$ were collected from the region of interest
in the 0.5--7 keV band.  We used \textsc{CIAO} 
(v.~2.2; Elvis et al.\ 2002, in prep.) and our own \textsc{IDL} routines
to prepare the data to the imaging and spectral studies.
The X-ray center was fixed to the peak of the projected mass from weak
lensing analysis (Lombardi et al.\ 2000) at (RA, Dec; 2000)${} =
(10^\mathrm{h} 10^\mathrm{m} 32\fs68, -12^{\circ} 39' 58.8")$.  Note
that the maximum value in a $5\arcsec$-smoothed image of the cluster
X-ray emission is at (RA, Dec)${} = (10^\mathrm{h} 10^\mathrm{m} 32\fs44,
-12^{\circ} 39' 55.6")$, i.e. less than 5 arcsec apart from the adopted
center.  With respect to the adopted center, a clear
asymmetry in the surface brightness distribution is however detected,
suggesting an excess in emission in the northern region (see
Fig.~\ref{xray_asi}). 

\begin{figure}
  \begin{center}
    \epsfig{file=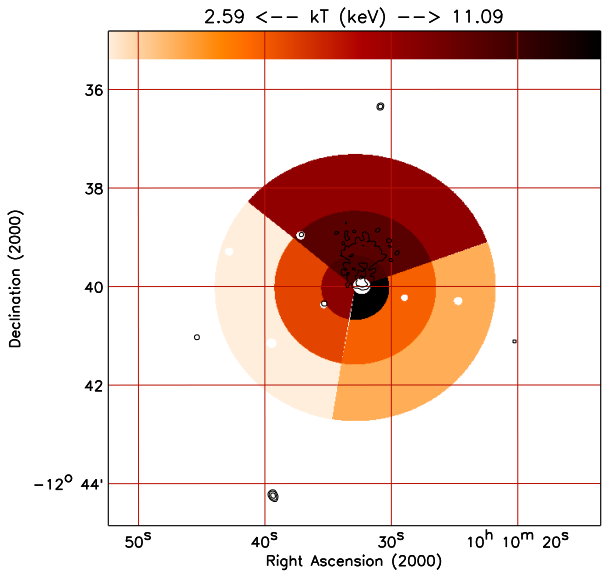,width=0.45\textwidth}
    \caption{Map of the distribution of the gas temperature according
      to the structures present in the X-ray surface brightness (see
      Fig.~\ref{xray_asi} and Sect.2.1).
      The code color represents also the variation in the range
      $(-3.1, +2.1) \sigma$ with respect to the global value of $7.2
      \mbox{ keV}$.}
    \label{fig:mapt}
  \end{center}
\end{figure}

\subsection{X-ray analysis}

\begin{figure*}
  \begin{center}
    \hbox{
      \epsfig{file=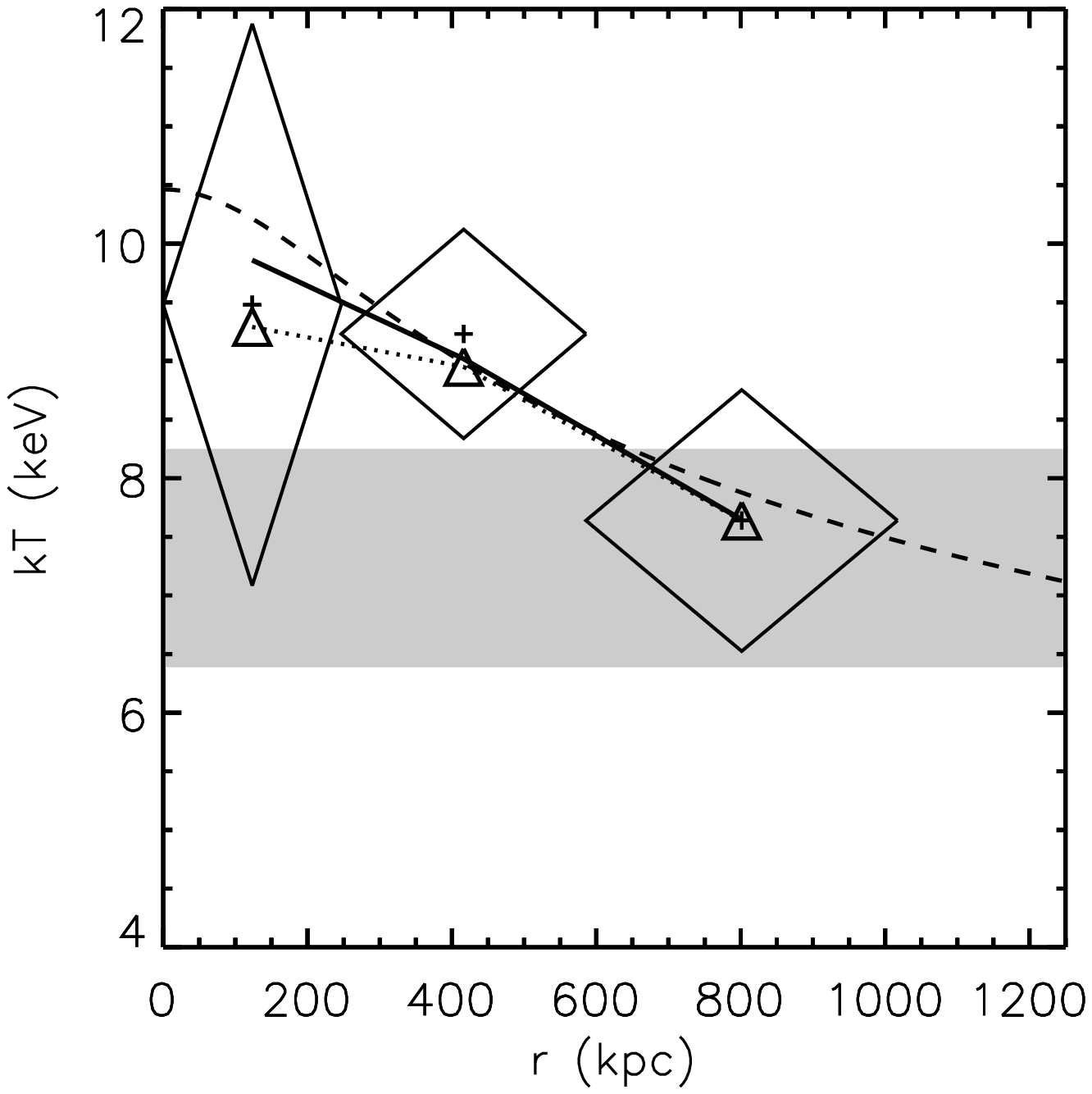,width=0.48\textwidth}
      \epsfig{file=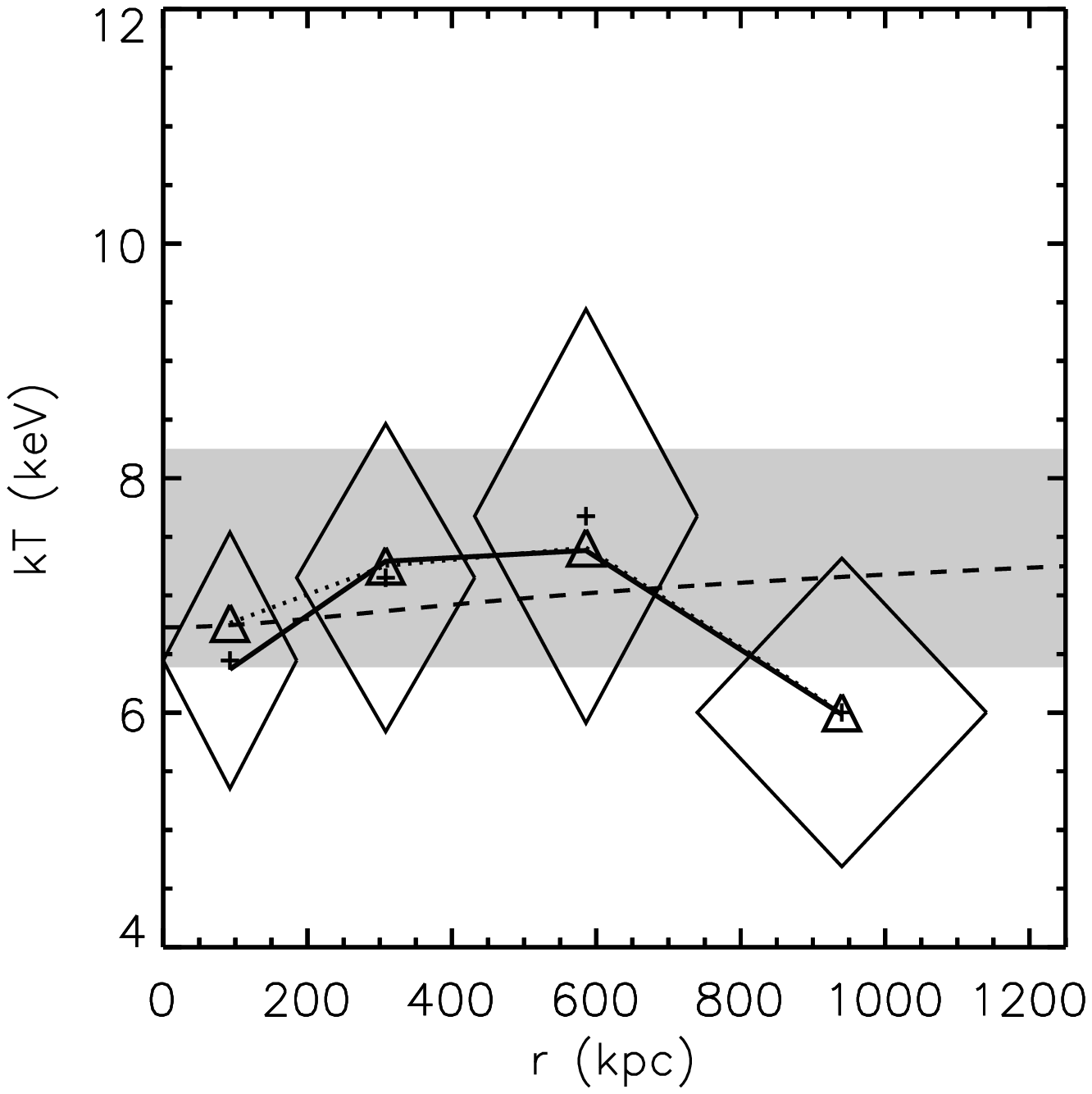,width=0.48\textwidth}
    }
    \caption{ Projected (triangles, dotted line) and deprojected
      (diamonds) gas temperature profiles.  The solid line shows the
      best-fit applying the hydrostatic equation in combination with
      the deprojected electron density as discussed in Sect.~2.2.  The
      dashed line indicates the best-fit from a polytropic model.
      Results for the North excess (left) and for azimuthally averaged
      profile (right) are shown with overplotted the $1 \sigma$ range
      of emission-weighted temperature up to $110 \mbox{ kpc}$.}
    \label{fig:temp} \end{center}
\end{figure*}

We detected extended emission at $2 \sigma$ confidence level 
up to 4.1 arcmin ($1.55 \mbox{ Mpc}$) and we were able to extract
a total of four spectra with about $2\,000$ source counts each to put
reasonable constraints on the plasma temperature profile up to $1140
\mbox{ kpc}$.  The contribution from the source to the total count rate
decreases from more $95\%$ in the innermost spectrum to about $30\%$
in the outermost one.  We also evaluated the distribution of the
plasma temperature values in sectors according to the asymmetrical
surface brightness shown in Fig.~\ref{xray_asi}.  These best-fit
values for a two-dimensional map and for an azimuthally averaged
profile are represented in Figs.~\ref{fig:mapt} and \ref{fig:temp}.

We obtained the 
Redistribution Matrix Files (RMFs) and Auxiliary Response Files (ARFs)
by using the \textsc{CIAO} routines \texttt{mkrmf} and \texttt{mkarf}
with the QEU files included in the \textsc{CtiCorrector}
package.
An emission from an optically-thin plasma (\textsc{Mekal}~--~Kaastra
1992, Liedhal et al.\ 1995, in \textsc{XSPEC} v.~11.1.0~--~Arnaud
1996) with the metal abundance fixed to 0.3 times the solar value
(Anders \& Grevesse 1989) and absorbed from the interstellar medium
parametrized using the T\"ubingen-Boulder model (\texttt{tbabs} in
\textsc{XSPEC}; Wilms, Allen \& McCray 2000) was adopted to reproduce
the observed spectra.  A galactic column density fixed to 
$7.0 \times 10^{20} \mbox{ cm}^{-2}$ (from radio HI maps in Dickey \&
Lockman 1990) was assumed.  A local background was adopted also
considering the relatively high column density of this field with
respect to the blank field available for the same CCD and the proper
observational period.
The overall spectral fit of the counts collected within $1100 \mbox{
kpc}$ from the adopted center provides an emission weighted
temperature of $7.2^{+1.0}_{-0.8} \mbox{ keV}$ and a bolometric
luminosity of $1.6 \times 10^{45} \mbox{ erg s}^{-1}$ ($0.9 \times
10^{45}$ in the $2$--$10 \mbox{ keV}$ band).

\subsection{X-ray mass profile}

In accordance with the weak lensing analysis described in the following
section, we assume a spherical geometry for
both the dark matter halo and the X-ray emitting plasma 
(note that negligible effects, when compared with our statistical 
uncertainties, can be introduced on the mass estimate due
to the aspherical X-ray emission, see, e.g., Piffaretti 
et al. 2002). 
The values of gas density and temperature in volume shells are
recovered from the projected spectral results as described 
in Ettori et al.\ (2002).
To measure the total gravitating mass $M_\mathrm{tot}$, we then
constrained the parameters of an assumed mass model by fitting the
deprojected gas temperature (shown in Fig.~\ref{fig:temp}) with the
temperature profile obtained by inversion of the equation of the
hydrostatic equilibrium between the dark matter potential and the
intracluster plasma, 
\begin{eqnarray}
-G \mu m_{\rm p} \frac{n_{\rm e} M_{\rm tot, model}(<r)}{r^2} =
\frac{d\left(n_{\rm e} \ kT\right)}{dr} \; .
\label{eq:mtot}
\end{eqnarray}
In this equation, $\mu=0.6$ is the mean molecular weight in a.m.u.,
$G$ is the gravitational constant, $m_{\rm p}$ is the proton mass, and
$n_{\rm e}$ is the deprojected electron density.  
We considered both the King approximation to the isothermal sphere (King 1962) 
and a Navarro, Frenk \& White (1997) dark matter density profile
as mass models (see details in Ettori et al.\ 2002). 
By fitting the temperature profile in Fig.~\ref{fig:temp}, we measured the
best fit parameters $(r_{\rm s}, c) = (646 \pm 390, 4.2 \pm 0.9)$ for
a King and $(1122 \pm 287, 2.3 \pm 0.8)$ for a NFW mass model.
 

Using a $\beta$-model (Cavaliere \& Fusco-Femiano 1976) and the
\textit{ROSAT\/} HRI surface brightness profile detected up to about
$765 \mbox{ kpc}$ at the $2 \sigma$ level, Lewis et al.\ (1999)
measured $M_{\rm tot} = (3.8 \pm 0.6) \times 10^{14} M_{\odot}$,
assuming an isothermal gas temperature of $7.3 \mbox{ keV}$.  If we
apply a $\beta$-model to our surface brightness profile and a
polytropic function to the gas temperature profile, we obtain
$r_\mathrm{c} = 0.264^{+0.025}_{-0.022} \mbox{ Mpc}$, $\beta =
0.618^{+0.025}_{-0.031}$ and $\gamma = 0.96^{+0.15}_{-0.16}$.
The derived mass estimate is lower by $10$--$20\%$ 
(by $12\%$ at $765 \mbox{ kpc}$) than the one in Lewis et al. (1999).
The best-fit mass models give $M_\mathrm{tot, King} = (4.5 \pm 1.2)
\times 10^{14} M_{\odot}$ and $M_\mathrm{tot, NFW} = (3.2 \pm 0.5)
\times 10^{14} M_{\odot}$, in agreement with the results obtained from
each independent sector of the temperature map in Fig.~\ref{fig:mapt}
[e.g., the region to North, which has a higher signal-to-noise ratio
due to the excess in brightness, gives $M_{\rm tot, King} = (4.5 \pm
0.8) \times 10^{14} M_{\odot}$ and $M_{\rm tot, NFW} = (4.1 \pm 0.6)
\times 10^{14} M_{\odot}$].

\begin{figure*}
\begin{center}
\hbox{
\epsfig{file=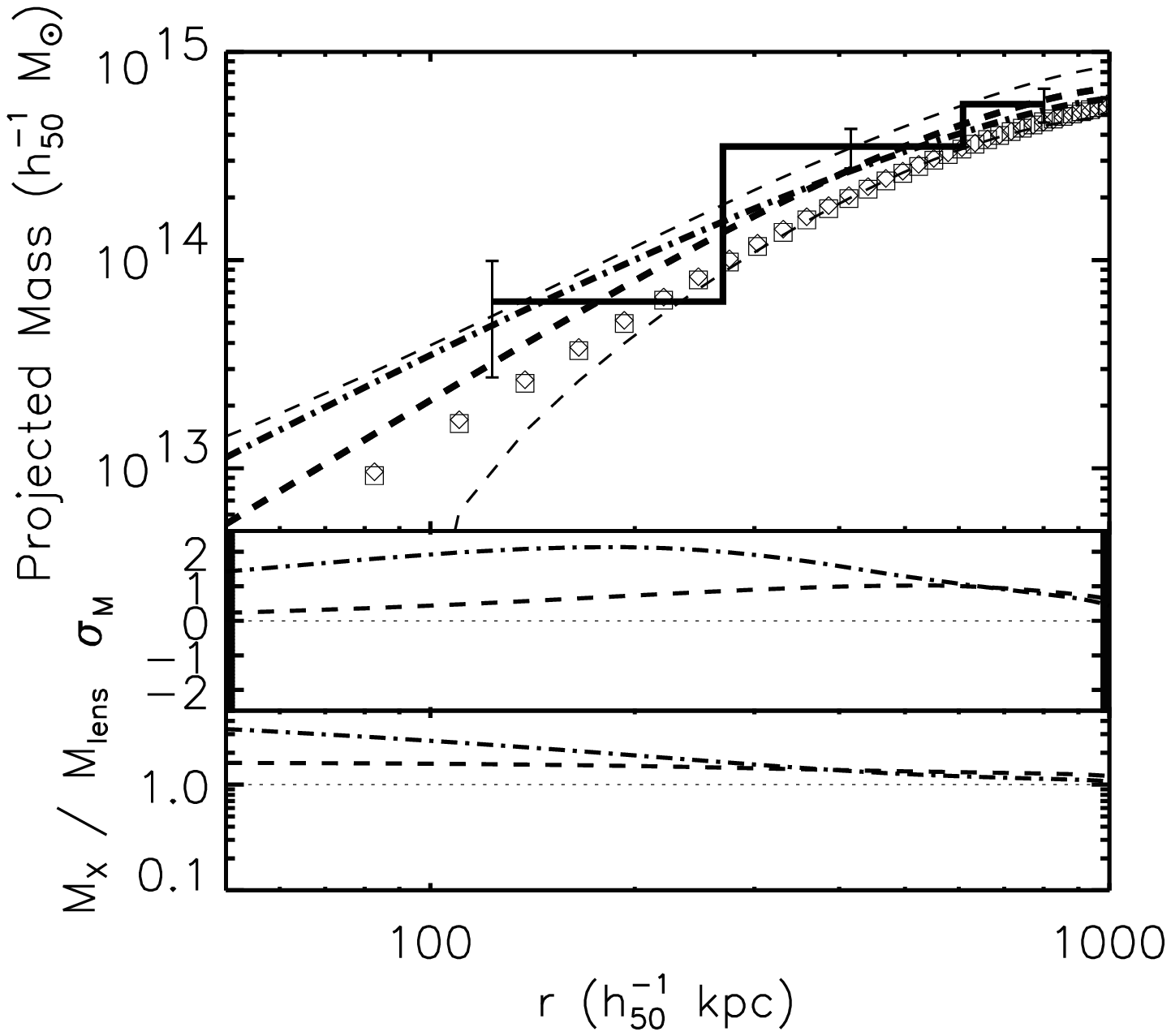,width=0.48\textwidth}
\epsfig{file=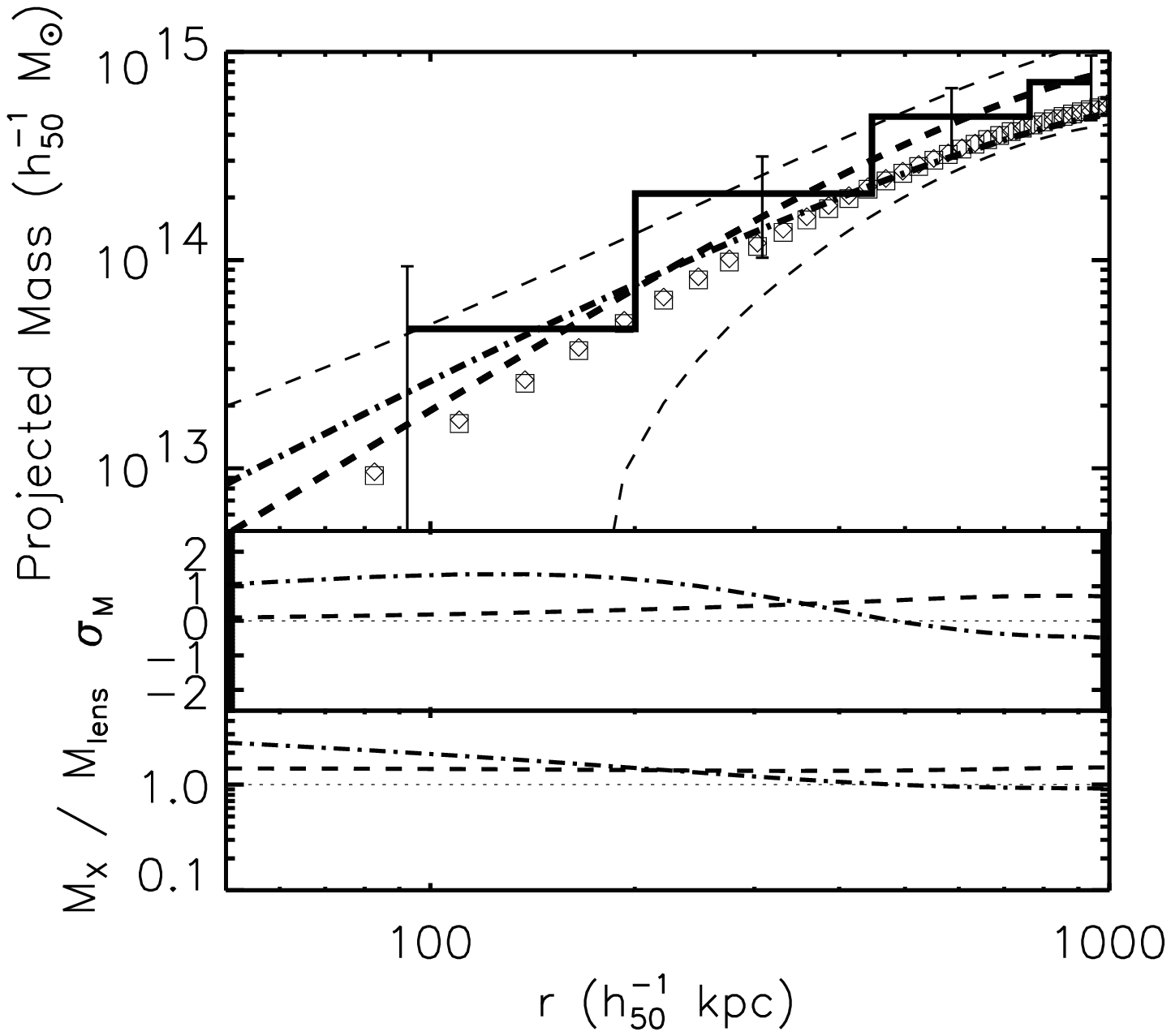,width=0.48\textwidth}
} \caption{
Projected X-ray total gravitating mass profile (with relative
error) obtained directly from the electron density and temperature
profiles is shown as histogram.
Cumulative projected mass profiles from X-ray best-fit mass model
(King: {\it dashed line}; NFW: {\it dot-dashed line})
and weak lensing analysis ({\it squares}/B band, {\it diamonds}/V band)
are overplotted.
The deviations in $\sigma$ and in the value of the ratio between
X-ray best-fit mass model and weak lensing projected masses are
plotted in the panels at the bottom.
} \label{fig:mass} \end{center}
\end{figure*}

\section{Weak lensing mass}
\label{sec:weak-lens-analys}

A weak lensing analysis of MS1008.1$-$1224 using VLT-FORS1 images was
carried out by Lombardi et al. (2000) and is summarized below. A
parallel weak lensing analysis carried out by Athreya et al. (2002),
leads to a mass estimates in agreement with the one presented here.

\subsection{Weak lensing analysis}
\label{sec:weak-lens-analys-1}

The weak lensing analysis was performed \textit{independently\/} on
the four (B, V, R, and I) FORS-1 optical images using the
\textsc{IMCAT} package (Kaiser et al. 1995; see also Kaiser \& Squires
1993).  After generating source catalogs, we separated stars from
galaxies; the measured sizes and ellipticities of stars were used to
correct the galaxy ellipticities for the PSF (see Kaiser et al. 1995;
Luppino and Kaiser 1997).  The observed galaxies were classified as
foreground, background, and cluster members using, when available, the
redshift from the CNOC survey (Yee et al. 1998), or the observed
colors otherwise.  Finally, we used the fiducial background galaxies
to obtain the shear field of the cluster and, from this, the projected
mass distribution (see Lombardi \& Bertin 1999).

Our study differs on a few points with respect to similar weak lensing
analyses:
\begin{itemize}
\item We decided to use a robust, median estimator to obtain the local
  shear map from the galaxy ellipticities instead of the more common
  simple average.
In particular, we estimated the local shear on each point of the map
by taking a (weighted) median on the observed ellipticities of
angularly close background galaxies (see Sect.~4.4 of Lombardi et
al. 2000).  This way we make sure that a few galaxies with poorly
determined ellipticities do not significantly affect the shear
estimation.

\item We estimated the background sources redshift distribution by
  resampling the catalog of photometric redshifts of Fern\'andez-Soto
  et al. (1999) in the Hubble Deep Field.  

\item We took advantage of the multi-band observations by performing
  the weak lensing analyses on each band separately.
\end{itemize}

\subsection{Weak lensing mass profile}
\label{sec:results}

The reconstructed two-dimensional mass distribution appears to be 
centered on the cD galaxy and shows elliptical profiles oriented in
direction North-South.  No substructure on scales larger
than $30''$ was detected.
In order to remove the \textit{mass-sheet degeneracy} (see, e.g., 
Kaiser \& Squires 1993), we fitted the mass profiles 
with non-singular isothermal sphere models; 
from this we obtain, for example, $M(r < 1 \, h_{50}^{-1} \mbox{ Mpc}) 
= 5.3 \times 10^{14} \, h_{50}^{-1} \mbox{M}_\odot$.  
We note that, because of the smoothing operated in the
two-dimensional mass maps, the weak lensing mass profile for $r < 1'$
is bound to be underestimate.  On the other hand, at large radii (say,
$r > 3'$), an error on the removal of the mass-sheet degeneracy can in
principle lead to an unreliable ``total'' weak lensing mass estimate.
Note that the four profiles (from B, V, R and I optical images)
agree very well to each other, which strongly support the results 
of our analysis.

\section{Conclusions}

The differential X-ray best-fit mass model has been
weighted by the relative portion of the shell observed in each ring
and, then, cumulated up to the radius of 1020 kpc (the outer
radius of our spatially resolved spectroscopy).
In Fig.~\ref{fig:mass}, we plot and compare the projected mass profiles
of the galaxy cluster MS1008.1$-$1224 obtained from 
both the spatially resolved spectral analysis of the \chandra
observation and the weak lensing analysis of FORS1-VLT
multicolor imaging. The two independently reconstructed
mass profiles agree very well within $1 \sigma$ uncertainty
both in absolute values and in the overall shape of the 
profile. Note that the mass center is fixed to the peak of the 
lensing map density that is consistent with the X-ray peak
as discussed in Sect.~2. 
This result does not change when we consider the different 
density and temperature
profiles observed in the northern region where a significant
surface brightness excess is located and a higher signal-to-noise
ratio is available, arguing for the robustness of the X-ray mass
estimates once gas density and temperature distributions 
can be properly mapped.


\end{document}